\documentclass[12pt,a4paper,onecolumn]{article}
\usepackage[utf8]{inputenc}
\usepackage{amsmath}
\usepackage{amsfonts}
\usepackage{amssymb}
\usepackage{graphicx}
\usepackage{natbib}
\usepackage{ragged2e}
\usepackage{multirow}
\usepackage{color}

\begin{document}

\begin{center}
{\bf {\Large Magnetic fields of protoplanetary disks}}
\vspace{5mm}

S.A. Khaibrakhmanov$^{1\,2\,3}$
\vspace{5mm}

{\small $^1$Saint-Petersburg State University,
7-9 Universitetskaya Emb., St Petersburg 199034, Russia\\
$^2$Chelyabinsk State University,
129 Br. Kashirinykh St., Chelyabinsk 454001, Russia\\
$^3$Ural Federal University,
51 Lenina St., Ekaterinburg 620000, Russia}
\vspace{5mm}

e-mail: s.khaibrakhmanov@gmail.com
\end{center}

We review the current status of studies on accretion and protoplanetary disks of young stars with large-scale magnetic fields. Observational data on magnetic fields of the disks are compiled and analysed.  Modern analytical and numerical MHD models of protoplanetary disks are discussed. The mechanisms of angular momentum transport via turbulence, magnetic tensions and outflows are outlined. We consider the influence of Ohmic dissipation, magnetic ambipolar diffusion,  magnetic buoyancy, and the Hall effect on the evolution of the magnetic flux in disks. Modern MHD models of accretion disks show that the magnetic field can influence the structure of protoplanetary disks. We argue that the available observational data on the magnetic fields in protoplanetary disks can be interpreted within the framework of fossil magnetic field theory. We summarize the problems of the modern theory of accretion and protoplanetary disks with magnetic fields and also outline the prospects for further research.

\vspace{5mm}
{\it keywords}: accretion, accretion disks; magnetic fields, magnetohydrodynamics (MHD), protoplanetary disks

\section{Introduction}

Observations in infra-red (IR), optical, ultraviolet (UV) waves and X-rays (XR) indicate that young stars are surrounded by gas-dust accretion disks. IR-excess in the spectral energy distributions of young stellar objects (YSO) is a sign of the dusty disk around the star, while UV- and XR-excess points to the process of gas accretion onto the stellar surface~\citep[see review by][]{hartmann2016}. Optical and IR-imaging allows to conclude that the accretion disks are geometrically thin with disk's thickness-to-radius ratio of $\lesssim 0.1$.  During the period of $1-10$~Myr, mass accretion rate onto the young star decreases from $\sim 10^{-6}$ to $10^{-9}\,M_\odot$~yr$^{-1}$, and mass of the disk correspondingly decreases. Modern instruments with high-spatial and spectral resolution, like The Atacama Large Millimeter/submillimeter Array (ALMA), Very Large Telescope (VLT) and James Webb Space Telescope (JWST), allow to study structure of the accretion disks of young stars with unprecedented details. Such studies have found that substructures in the form of concentric rings and spirals are ubiquitous in the accretion disks~\citep{andrews2018}. Observed substructures can be interpreted as sings of ongoing planet formation process in disks. In recent years, direct evidences of the protoplanets forming in the disks appeared~\citep{boccaletti2020}. Therefore, one conclude that, during their evolution, accretion disks of young stars evolve into protoplanetary disks (hereafter PPDs), which are analogues of the protosolar nebula.

Elaborating a theory of planet formation requires developing models of accretion and protoplanetary disk evolution that take into account a variety of physical and chemical processes characterized by a wide range of spatial and temporal scales. One of the main components of the PPD models is the magnetic field. On the one hand, star formation theory predicts that the accretion disks of young stars are born with a large-scale fossil magnetic field, which is supported by contemporary observational data~\citep[see reviews by][]{mckee2007, dkh2015}. On the other hand, the efficiency of angular momentum transport in disks depends on the strength and geometry of the magnetic field~\citep[see reviews by][]{paplin1995, lesur2021}. 

In this paper, we compile and analyse observational data on the magnetic field strength and geometry in PPDs (section~\ref{sec:obs}), as well as review current state of the theory of the evolution of PPDs with magnetic field (section~\ref{sec:theory}). Specific attention is paid to the origin of the magnetic field (section~\ref{sec:origin}), role of MHD effects in magnetic field evolution (section~\ref{sec:mhd}) and mechanisms of angular momentum transport (section~\ref{sec:angmom}). Main results achieved in the field of MHD modelling of the accretion and protoplanetary disks are also presented (section~\ref{sec:dynamics}). As a summary, we outline main problems and perspectives in both observations and theory (section~\ref{sec:summary}). The review focuses on low- and intermediate-mass young stars, namely T Tauri and Herbig Ae/Be stars, although some complementary information concerning young high-mass stars is also briefly discussed.

\section{Observations}
\label{sec:obs}

Let us consider pathways to study the magnetic fields in PPDs observationally.

\paragraph{Outflows and jets.}
Outflow phenomena are just as ubiquitous, as accretion in YSOs~\citep[see][]{frank2014}. Outflows can be divided into two groups: slow, wide-angle molecular outflows with speeds up to $30$~km/s, and fast collimated molecular and/or atomic jets with speeds of $100-1000$~km/s. In many cases fast jets are enclosed in slow outflows~\citep[e.g.][]{mundt1983, arce2002}. In such cases, the wide-angle outflow can be interpreted as a slowly expanding cavity produced by the propagation of the fast jet into the interstellar medium~\citep[see][]{bally2016}. The outflows are usually bipolar, but there are also examples of one-sided flows~\citep{codella2014}. 
Several mechanisms have been proposed for the formation of the outflows, each  requiring the presence of a magnetic field of specific strength and geometry in the disk~\citep[e.g.,][see also next section]{pudritz2019}. According to numerical simulations, fast collimated jets probably form in the disk regions closest to the central young star, while slower outflows emanate from the disk at greater distances~\citep{machida2008}. Therefore, outflows and jets are indirect evidence of the presence of the magnetic field in accretion disks and PPDs. Synchrotron radiation in protostellar jets of high-mass protostars indicates the presence of the magnetic field in the jet, while linear polarization in molecular lines (Goldreich-Kylafis effect) allows to probe the magnetic field geometry and strength in the jets of low-mass protostars~\citep{lee2020}.

\paragraph{Masers.}
The region of jet interaction with interstellar medium is characterized by the formation of multiple shocks. The conditions in the post-shock region may be favourable for the formation of class I methanol masers, OH and H$_2$O masers~\citep[e.g.,][]{elitzur1989, sobolev2005, voronkov2006}, which are therefore considered as an instrument for studying the outflows in both low- and high-mass YSOs~\citep[e.g.,][]{gomez1995, furuya2003, moscadelli2006, voronkov2006, kalenskii2013, ladeyschikov2022}. 
Masers trace the gas kinematics in the outflow, which can be used to test MHD models of the outflows and draw conclusions about the magnetic field in the region of the outflow~\citep{moscadelli2022}. Measurements of the Zeeman splitting in OH and H$_2$O maser lines give information about magnetic field strength in the outflow~\citep{sarma2002, vlemmings2008, goddi2017}.

\paragraph{Polarization mapping.}
Polarization mapping allows to study the magnetic field geometry in the interstellar medium. 

The first polarization studies of the regions near T Tauri stars have found evidence of the large-scale magnetic field aligned perpendicular to the disk planes~\citep{tamura1989}. Such a magnetic field can be attributed to the outflows from the disks. The derived magnetic field direction corresponds to the large-scale magnetic field of the surrounding molecular cloud, which is an evidence that the magnetic field of YSOs is genetically linked to the magnetic field of the parent molecular clouds.

The ALMA interferometer's capability enables to make spatially resolved polarization maps of PPDs and study their magnetic fields~\citep{bertrang2017, brauer2017}. The polarization mapping of several disks around T~Tauri and Herbig Ae/Be stars with spatial resolution of $50$~au has revealed magnetic field with complex geometry in the disks~\citep{li2016, li2018}. Interpretation of the polarization maps is complicated due to the effects of self-scattering and radiation anisotropy. The maps of the polarization emission at wavelengths $\lambda\sim 1$~mm are well described by the models taking into account the self-scattering by the dust grains of maximum size $a_{\rm max}\sim \lambda/2\pi$~\citep{kataoka2015}.

\paragraph{Zeeman effect.}
The magnetic field strength in YSOs can be studied using Zeeman effect measurements in different emission lines~\citep[e.g.,][]{lankhaar2023}, as well the Chandrasekhar-Fermi method. The observed Zeeman broadening of the optical emission lines in the spectra of T Tauri stars corresponds to the magnetic fields of $1-3$~kG~\citep{krull2007, yang2011}. \cite{donati2005} reported the first detection of the Zeeman broadening of the emission lines from the inner region of FU Ori accretion disk, indicating the presence of a magnetic field with strength of $1$~kG at the radial distance $r=0.05$~au from the star. Zeeman splitting measurements of CN molecular lines are another promising tool for estimating the magnetic field strength. \cite{vlemmings2019} attempted to detect the Zeeman effect in the CN emission from the TW~Hya disk. The authors found no signs of the Zeeman splitting, which sets an upper limit on the magnetic field strength of $8\cdot 10^{-4}$~G at a distance $42$~au from the star. \citet{harrison2021} used a similar approach and found upper limits of $0.087$~G for the toroidal magnetic field and of $0.061$~G for the vertical magnetic field in the disk AS 209.

\paragraph{Remnant magnetism of the solar system meteorites.}
Asteroids and comets in the Solar system are made of material and/or contain inclusions that were formed during the formation of the solar system. Therefore, measurements of the remnant magnetisation of meteorites as well as small bodies in the Solar system give indirect estimates of the magnetic field strength in the protosolar nebula. The study of meteorites of various types shows that the magnetic field in the region of $r=1-7$~au can vary in range of $0.02-1$~G. Comparing the formation time of different meteorite inclusions allows to speculate about the evolution of the magnetic field in the region of terrestrial planet formation in the protosolar nebula~\citep{borlina2022}.

Moreover, measurements with the Rosetta Magnetometer and Plasma
Monitor have shown that comet 67P/Churyumov-Gerasimenko (67P) has a surface magnetic field of less than $0.03$~G, which could be attributed to the magnetic field in the protosolar nebula in the region of $15-45$~au from the protosun~\citep{biersteker2019}.

\begin{table}[htb!]
{\footnotesize
\caption{Observational constraints on the magnetic field strength in YSOs}

\begin{center}
\begin{tabular}{p{0.94in}p{0.94in}p{0.8in}p{0.94in}p{1.08in}}
\hline
\hline
  \Centering{object} &   \Centering{$B$, G} &   \Centering{$r$, au} &   \Centering{method} &   \Centering{refs} \\
\hline
\multicolumn{5}{c}{{\it Young stars}}\\
\Centering{Taurus/Auriga region} & \Centering{$(1.12 - 2.9)\cdot 10^3$} & \Centering{$\sim 0.01$} & \Centering{Zeeman broadening in Ti lines} & \Centering{\cite{krull2007}} \\
\Centering{Orion
Nebula Cluster} & \Centering{$(1.3 - 3.14)\cdot 10^3$} & \Centering{$\sim 0.01$} & \Centering{-//-} & \Centering{\cite{yang2011}} \\
\Centering{Herbig Ae/Be stars} & \Centering{$37 - 222$} & \Centering{$\sim 0.01$} & \Centering{Zeeman splitting} & \Centering{\cite{hol2019}} \\
\hline
\multicolumn{5}{c}{{\it Protosolar nebula}}\\
\Centering{ordinary chondrites} & \Centering{$0.54\pm 0.21$} & \Centering{$1-3$} & \Centering{remnant magnetism} & \Centering{\cite{fu2014}} \\
\Centering{volcanic angrites} & \Centering{$<0.06$} & \Centering{$2-3$} &\Centering{-//-} & \Centering{\cite{wang2017}} \\
\Centering{carbonaceous chondrites} & \Centering{$1.08\pm 0.15$} & \Centering{$3-7$} & \Centering{-//-} & \Centering{\cite{butler1972}} \\
\Centering{-//-} & \Centering{$0.021\pm 0.015$} &  \Centering{-//-} & \Centering{-//-} & \Centering{\cite{cournede2015}}\\ 
\Centering{-//-} & \Centering{$<0.08\pm 0.043$} &  \Centering{-//-} & \Centering{-//-} & \Centering{\cite{fu2020}}\\
\Centering{-//-} & \Centering{$1.01\pm 0.48$} &  \Centering{-//-} & \Centering{-//-} & \Centering{\cite{borlina2021}}\\
\Centering{-//-} & \Centering{$>0.4$} &  \Centering{-//-} & \Centering{-//-} & \Centering{\cite{fu2021}}\\
\Centering{-//-} & \Centering{$<0.009$}&  \Centering{-//-} & \Centering{-//-} & \Centering{\cite{borlina2022}}\\
\Centering{Comet 67P} & \Centering{$<0.03$} & \Centering{$15-45$} & \Centering{direct} & \Centering{\cite{biersteker2019}} \\
\hline
\multicolumn{5}{c}{{\it protoplanetary disks}}\\
\Centering{Fu Ori} & \Centering{$1000$} & \Centering{$0.05$} & \Centering{Zeeman broadening} & \Centering{\cite{donati2005}} \\
\Centering{TW Hya} & \Centering{$<8\cdot 10^{-4}$} & \Centering{$42$} & \Centering{Zeeman in CN} & \Centering{\cite{vlemmings2019}} \\
\hline
\label{Table:B}
\end{tabular}
\end{center}
}
\end{table}

\paragraph{Analysis of the observational data}
Data on the magnetic field strength in PPDs are compiled in Table~\ref{Table:B}.  The table lists object name (column~1), magnetic field strength (column~2), corresponding distance from the star (column~3), method of measurement (column~4), and references (column~5). 

Although the data collected correspond to different objects, some general conclusions can be made based on table 1. The data indicate that the magnetic field strength in PPDs decreases with distance from the star raging from $\sim 1$~kG near the stellar surface to $10^{-4}-10^{-3}$ at several tens of au.%

\section{Theory}
\label{sec:theory}
\subsection{Origin of the magnetic field in disks}
\label{sec:origin}
Analysis of the observational data presented in section~\ref{sec:obs} leads to the conclusion that accretion and protoplanetary disks of young stars have a large-scale magnetic field. The origin of this magnetic field is naturally explained within the framework of modern theory of star formation. 
According to the theory, stars form as a result of gravitational collapse of protostellar clouds, which are cold, dense cores of interstellar molecular clouds~\citep[see review by][]{mckee2007}. ISM polarization mapping and Zeeman splitting measurements of OH lines in protostellar clouds indicate that the clouds are threaded by a large-scale galactic magnetic field with typical magnetic field strength of $10^{-4}-10^{-5}$~\citep[see recent review by][]{crutcher2019}. 
Numerical simulations of star formation in the protostellar clouds have shown that the magnetic flux of the clouds is partially preserved during star formation. Therefore, the magnetic field of young stars and their accretion disks is of {\it fossil} nature~\citep{dud1995, dkh2015}.

Cyclonic turbulence or convection (so-called $\alpha$-effect) together with differential rotation ($\Omega$-effect) may lead to the dynamo-generation of the magnetic field from a small seed field. The dynamo-generation of the magnetic field supports magnetic field of T Tauri stars after the onset of convection in star interior~\citep{dud1995}. Hydromagnetic dynamo could also operate in turbulent accretion and protoplanetary disks under certain conditions~\citep{brandenburg1995, gressel2015b, moss2016}. The fossil magnetic field can be a seed field for the dynamo.

\subsection{MHD effects}
\label{sec:mhd}

In order to analyse influence of the magnetic field on the dynamics of PPDs, let us consider following system of MHD equations:
\begin{eqnarray}
  \frac{\partial\rho}{\partial t} + {\bf \nabla}\cdot\left(\rho\textbf{v}\right) &=& 0, \label{Eq:Cont}\\
  \rho\frac{d {\bf v}}{dt} &=& -\nabla \left(p + \frac{B^2}{8\pi}\right) + \rho{\bf g} + \frac{1}{4\pi}\left({\bf B}\cdot \nabla\right){\bf B} + \mathrm{div}\hat{\sigma}',\label{Eq:Motion}\\
  \rho T\frac{ds}{dt} &=& \sigma_{ik}^{\prime}\frac{\partial v_i}{\partial x_k} + {\bf \nabla}\cdot \mathcal{\textbf{F}} + \Gamma,\label{Eq:Energy}\\
  \frac{\partial\textbf{B}}{\partial t} &=& \nabla\times \left((\textbf{v} + \textbf{v}_{\rm AD}) \times \textbf{B}\right) + \nabla\times \left(\eta_{\rm OD}\nabla\times \textbf{B}\right) + \nonumber \\
  & & \nabla\times\left(\frac{c}{4\pi en_{\rm e}}\left((\nabla\times\textbf{B})\times\textbf{B}\right)\right) + \nabla\times \left(\textbf{v}_{\rm b} \times \textbf{B}_{\rm t}\right). \label{Eq:Induction}
\end{eqnarray}

Equation of motion (\ref{Eq:Motion}) includes gas and magnetic pressure gradients, stellar gravity $\textbf{g}$, magnetic tensions (the third term on the right-hand-side) and turbulent stresses described by the stress tensor $\hat{\sigma}'$ ($\sigma_{ik}^{\prime}$ in the index notation, indices $i$ and $k$ number the spatial directions, see~\citet{LL8}). The turbulent stresses are treated following~\citet{ss1973}, who suggested to replace the viscous stress tensor with the turbulent one. This assumption  reflects the fact that the turbulence acts as an anomalous viscosity. Accurate justification of this idea has been done later in the works on the turbulence driven by the magnetorotational instability in disks~\citep[see][]{balbus1998, shakura2018}. Turbulent stress tensor in the model of \citet{ss1973} is estimated as: $\sigma'_{r\varphi}=-\alpha p$, where non-dimensional parameter $\alpha\leq 1$ describes the efficiency of turbulence. Corresponding kinematic turbulent viscosity is $\nu_{\rm t}=\alpha c_{\rm s}H$, where $c_{\rm s}$ is the sound speed, $H=c_{\rm s}/\Omega_{\rm k}$ is the scale height of the disk, $\Omega_{\rm k}$ is the keplerian angular frequency. It should be noted that, generally speaking, small-scale magnetic field tensions are also incorporated into $\alpha$.  Mean accretion rates observed in PPDs are of $10^{-8}\,M_\odot/\mbox{yr}$, which corresponds to $\alpha \approx 0.01$~\citep[see, e.g.,][]{hartmann2016}.

Equation of heat transfer (\ref{Eq:Energy}) describes the thermal structure of the disk with temperature $T$ and specific entropy $s$. The first term on the right-hand-side of (\ref{Eq:Energy}) represents gas heating due to dissipation of turbulence, the second term describes the heat losses due to radiation energy flux from the disk $\mathcal{\textbf{F}}$. In the optically thick case~\citep[e.g., ][]{zeldovich_BOOK, shakura2018},
\begin{equation}
\mathcal{\textbf{F}} = -\frac{c}{3\chi_{\rm R}}\nabla\epsilon,
\end{equation}
where $\chi_{\rm R} = \kappa_{\rm R}\rho$, $\kappa_{\rm R}$ is the Rosseland mean opacity, $\epsilon = aT^4$ is the radiation energy density, $a$ is the radiation constant. Value $\Gamma$ accounts for additional heating sources, for example, Ohmic and ambipolar heating.

Induction equation (\ref{Eq:Induction}) is written taking into account magnetic ambipolar diffusion with speed $\textbf{v}_{\rm AD}$, Ohmic dissipation characterized by diffusivity $\eta_{\rm OD}$ (also known as magnetic viscosity), the Hall effect (the third term on the right-hand-side), and magnetic buoyancy characterized by buoyancy speed~$\textbf{v}_{\rm B}$.

Stationary speed of magnetic ambipolar diffusion is determined as~\citep[see][]{spitzer_BOOK, dud1989},
\begin{equation}
\textbf{v}_{\rm AD} = \frac{\left(\nabla\times \textbf{B}\right)\times \textbf{B}}{4\pi R_{\rm in}},
\end{equation}
where $R_{\rm in}=\mu_{\rm in}n_{\rm i}n_{\rm n}\langle\sigma v\rangle_{\rm in}$ is the the coefficient of friction between ions and neutrals, $\mu_{\rm in}$ is the reduced mass for ion and neutral particles, $n_{\rm i}$ and $n_{\rm n}$ are the ion and neutral number densities respectively, $\langle\sigma v\rangle_{\rm in}=2\times 10^{-9}$~cm$^3$~s$^{-1}$ is the coefficient of momentum transfer in collisions between ions and neutrals. In the linearised case, magnetic ambipolar diffusion can be described with a diffusivity
\begin{equation}
\eta_{\rm AD} = \frac{B^2}{4\pi R_{\rm in}}.
\end{equation}
Ambipolar diffusion leads to plasma heating at the volume rate $\Gamma_{\rm AD} = \dfrac{1}{2}v_{\rm AD}^2R_{\rm in}$ \citep{scalo1977}.

Ohmic diffusivity is determined as~\citep[e.g,][]{alfven_BOOK}
\begin{equation}
\eta_{\rm OD} = \frac{c^2}{4\pi\sigma_{\rm e}},
\end{equation} 
where
\begin{equation}
\sigma_{\rm e} = \frac{e^2n_{\rm e}}{m_{\rm e}\nu_{\rm en}},
\end{equation}
is the conductivity, $e$ is the electron charge,  $n_{\rm e}$ is the electrons number density, $m_{\rm e}$ is the mass of the electron, and $\nu_{\rm en} = \langle\sigma v\rangle_{\rm en}n_{\rm n}$ is the mean collision rate between electrons and neutral particles. The rate of momentum transfer in electron-neutral collisions, $\langle\sigma v\rangle_{\rm en}$, can determined by fitting experimental data~\citep[see Appendix in][]{sano2000}. Ohmic dissipation heats the plasma at the volume rate $\Gamma_{\rm OD} = \dfrac{\eta_{\rm OD}}{4\pi} \left(\mathrm{rot}\textbf{B}\right)^2$.

The Hall effect in the linearised case can be described with the coefficient
\begin{equation}
\eta_{\rm H} = \frac{cB}{4\pi en_{\rm e}}.
\end{equation}

Magnetic buoyancy is introduced via magnetic buoyancy velocity $\textbf{v}_{\rm B}$. It is considered that magnetic buoyancy instability causes toroidal magnetic field $\textbf{B}_{\rm t}=\{0,\, B_\varphi,\, 0\}$ to split onto separate magnetic flux tubes~\citep{parker_BOOK}. The flux tubes rise from the region of runaway magnetic filed generation and carry away the excess of magnetic flux. The buoyancy speed can be estimated by equating the buoyancy and drag forces, which gives $v_{\rm B}\approx v_{\rm A}$, where $v_{\rm A}$ is the Alfv{\'e}n speed~\citep[see also][]{kh2017b}. Magnetic buoyancy in the accretion disks can be represented by the diffusivity $\eta_{\rm B}=v_{\rm B}H$~\citep{campbell_BOOK}.

Several non-dimensional parameters are usually introduced to characterize the relative role of various MHD effects.

\begin{itemize}

\item Magnetic Reynolds number:
\begin{equation}
R_{\rm m} = \frac{v_0 L_0}{\eta},
\end{equation}
where $v_0$ is the typical gas speed, $L_0$ is the typical spatial scale, $\eta$ is either Ohmic, or ambipolar diffusivity. The magnetic Reynolds number is a ratio of the induction term to the diffusion term in the induction equation~(\ref{Eq:Induction}). The case $R_{\rm m} \gg 1$ corresponds to perfect conductivity. In this case, the magnetic flux is conserved, and the magnetic field is tightly coupled to the plasma flow ({\it frozen} in plasma). This regime is commonly referred to as an {\it ideal} MHD. In the opposite case, $R_{\rm m} < 1$, a small conductivity leads to a violation of the flux freezing condition and a dissipation of the magnetic flux. Under such conditions, Ohmic dissipation and magnetic ambipolar diffusion operate ({\it non-ideal} MHD regime).

\item Plasma beta
\begin{equation}
\beta = \frac{8\pi p}{B^2}
\end{equation}
is the ratio of gas pressure to the magnetic pressure. Electromagnetic force influences the plasma flow dynamics in the case when $\beta \lesssim 1$. We will call further such case as {\it dynamically} strong magnetic field. In the opposite case, $\beta \gg 1$, magnetic field is {\it kinematic} one.

\item The Hall parameter (or magnetization parameter):
\begin{equation}
\beta_{\rm a} = \omega_{\rm a}\tau_{\rm a},
\end{equation}
where $\omega_{\rm a}$ is the gyrofrequency for particles of type `a', $\tau_{\rm a}$ is the typical time of the collisions between the particles of type `a' and other particles. The Hall parameters is the ratio of gyrofrequency to collision frequency and characterizes the degree of plasma anisotropy with respect to the magnetic field direction. In the case $\beta_{\rm a} \gg 1$ collisions between particles occur so frequently that the cyclotron rotation of charged particles ceases, and the plasma conductivity is isotropic. In the opposite case, each charged particle is tightly coupled to a specific magnetic field line, and the plasma conductivity is highly anisotropic.

\end{itemize}

Table~\ref{Table:MHD} lists main MHD effects and corresponding ranges of non-dimensional parameters. 

In the case of $R_{\rm } \gg 1$, the magnetic field is frozen in gas, and runaway generation of the magnetic field is possible. Under such conditions, a magnetic field can be generated up to the so-called {\it equipartition} value, which is characterized by plasma $\beta\sim 1$. In accretion and protoplanetary disks, such conditions prevail in the innermost region of the disk, where the ionization fraction and conductivity are high due to the thermal ionization of alkali metals. Magnetic buoyancy is a main mechanism limiting the runaway generation of the toroidal magnetic field in this region~\citep{dkh2014, kh2017b}. 

Ohmic dissipation and magnetic ambipolar diffusion operate in the limit of small magnetic Reynolds numbers. Ohmic dissipation prevails over ambipolar diffusion in the non-magnetized plasma, when the Hall parameters for both electrons ($\beta_{\rm e}$) and ions ($\beta_{\rm i}$) are small. Magnetic ambipolar diffusion is efficient in the opposite case of magnetized plasma, when $\beta_{\rm e}\gg 1$ and $\beta_{\rm i} \gg 1$. It consists in the joint drift of electrons and ions, which are tightly coupled to the magnetic field, through the neutral gas with drift velocity $\textbf{v}_{\rm AD}$. Collisions between charged and neutral particles moving relative to each other lead to the decay of the currents and the magnetic field dissipation. The Hall effect operates in the intermediate case, when electrons are magnetized  ($\beta_{\rm e}\gg 1$), while ions are not  ($\beta_{\rm i} \ll 1$). The Hall effect is the drift of plasma in the direction perpendicular both to the magnetic field and electric field. It corresponds to the case, when the electron gas is frozen in plasma.  It should be emphasized that Ohmic dissipation and ambipolar diffusion are dissipative MHD effects leading to magnetic flux decay and corresponding plasma heating. The Hall effect is non-dissipative one, which can only lead to transformation of magnetic field geometry. It can operate both in poorly and fully ionized plasma.

\begin{table}[t]
{\footnotesize
\caption{MHD effects and corresponding non-dimensional parameters}

\begin{center}
\begin{tabular}{ccc}
\hline 
effect & conditions \\ 
\hline 
\hline
flux freezing & $R_{\rm m}\gg1$ \\ 
Ohmic dissipation &  $R_{\rm m}<1$, $(\beta_{\rm e}, \beta_{\rm i})\ll1$ \\ 
magnetic ambipolar diffusion & $R_{\rm m}<1$, $(\beta_{\rm e}, \beta_{\rm i})\gg 1$ \\ 
magnetic buoyancy & $\beta\lesssim 1$ \\ 
the Hall effect & $\beta_{\rm e}\gg 1$, $\beta_{\rm i} \ll 1$ \\ 
\hline 
\label{Table:MHD}
\end{tabular} 
\end{center}

}
\end{table}

In order to compare the relative role of Ohmic dissipation, magnetic ambipolar diffusion, and the Hall effect in PPDs, let us write down the ratios of corresponding coefficients~\citep[see][]{kh2017}. Ratio of ambipolar to Ohmic diffusivities:
\begin{equation}
\frac{\eta_{\rm AD}}{\eta_{\rm OD}} = \beta_{\rm e}\beta_{\rm i}\frac{m_{\rm i} + m_{\rm n}}{m_{\rm n}} \propto \left(\frac{B}{n}\right)^2,\label{eq:A2O}
\end{equation} 
where $m_{\rm i}$ and $m_{\rm n}$ are the ion and neutral particle masses, respectively.

Ratio of the Hall coefficient to the Ohmic diffusivity:
\begin{equation}
\frac{\eta_{\rm H}}{\eta_{\rm OD}} = \beta_{\rm e} \propto \left(\frac{B}{n}\right),\label{eq:H2O}
\end{equation}
to the ambipolar diffusivity:
\begin{equation}
\frac{\eta_{\rm H}}{\eta_{\rm AD}} = \beta_{\rm i}^{-1}\frac{m_{\rm n}}{m_{\rm i} + m_{\rm n}} \propto \left(\frac{B}{n}\right)^{-1},\label{eq:H2A}
\end{equation}

\begin{figure}
\includegraphics[width=0.99\textwidth]{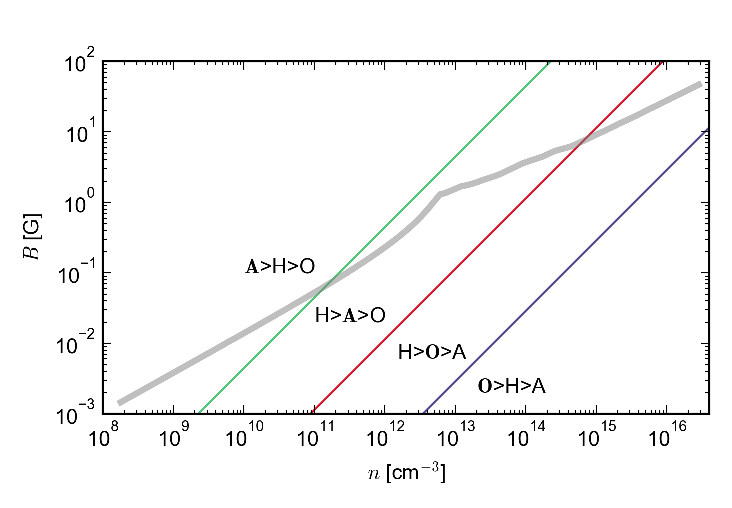}
\caption{Regions of dominant MHD effects in the $B-n$ plane (`A': magnetic ambipolar diffusion, `O': Ohmic dissipation, `H': the Hall effect). Thick line depicts dependence $B(n)$ predicted by the MHD model of~\citet{dkh2014} for typical parameters.}
\label{fig:MHD}
\end{figure}

Expressions (\ref{eq:A2O}--\ref{eq:H2A}) show that the relative role of the MHD effects depends on the magnetic field strength and the gas density. In Figure~\ref{fig:MHD}, we plot corresponding domains of dominant MHD effects in the $B-n$ plane. The ratios were evaluated following~\citet{kh2017} and using the approximate formula for the electron-neutral friction coefficient $\langle\sigma v\rangle_{\rm en}$ from \citet{sano2000}. Figure~\ref{fig:MHD} shows that Ohmic dissipation is the dominant dissipative MHD effect in the highest density regions (domains `H$>$O$>$A' and `O$>$H$>$A'). In contrast, magnetic ambipolar diffusion operates in low-density regions with strong magnetic fields (domains `H$>$O$>$A' and `O$>$H$>$A'). At the same time, the Hall effect operates in the domain of intermediate gas density (`H$>$A$>$O' and `H$>$O$>$A'). We emphasize that the Hall effect is not dissipative and in this sense can act {\it together} with Ohmic dissipation in region `H$>$O$>$A' and {\it together} with ambipolar diffusion in domain `H$>$A$>$O'. The Hall effect leads to the transformation of the magnetic field geometry in these domains, while magnetic diffusion causes magnetic flux dissipation and plasma heating.

We also plot dependence $B(n)$ in the accretion disk of the solar mass T~Tauri star calculated using the MHD model of~\citet{dkh2014} for typical parameters. Figure~\ref{fig:MHD} shows that all three MHD effects could operate in the accretion disks. Resulting strength and geometry of the magnetic field is discussed in Section~\ref{sec:dynamics} below. 

It should be noted that, generally speaking, turbulent diffusion has to be taken into account when considering the magnetic field evolution. Turbulent diffusivity $\eta_{\rm t}$ is connected to the turbulent viscosity via the turbulent magnetic Prandtl number,  $\mathrm{Pr}_{\rm m}=\nu_{\rm t} / \eta_{\rm t}$. The value of the magnetic Prandtl number in accretion discs is highly uncertain. Let us estimate  $\eta_{\rm t}$ for the case  $\mathrm{Pr}_{\rm m}=1$, which is considered a realistic value in MHD turbulence~\citep[see][and references therein]{guilet2014}. Using the typical gas temperature at $r=1$~au from the analytical solution of~\citet{dkh2014}, one obtains
\begin{equation}
\eta_{\rm t} = 7.6\times 10^9\left(\frac{\alpha}{0.01}\right)\left(\frac{T}{240\,\mathrm{K}}\right)\left(\frac{r}{1\,\mathrm{au}}\right)^{3/2}\left(\frac{M}{1\,M_\odot}\right)^{-1/2}\,\mathrm{cm}^2\,\mathrm{s}^{-1}.\label{eq:eta_turb}
\end{equation}

The turbulent diffusivity enters the induction equation linearly similar to Ohmic diffusivity. In the simplest case, the latter value depends on the ionization fraction, $x\approx n_{\rm e}/n_{\rm n}$, only. For $\langle\sigma v\rangle_{\rm en}\approx 10^{-8}$~cm$^2$~s$^{-1}$, we have
\begin{equation}
\eta_{\rm OD} = 2.8\times 10^{18}\left(\frac{x}{10^{-15}}\right)^{-1}\,\mathrm{cm}^2\,\mathrm{s}^{-1},\label{eq:eta_OD_est}
\end{equation}
where we used typical value $x(r=1\,\mathrm{au})\approx 10^{-15}$ from the analytical solution of~\citet{dkh2014}. The comparison of (\ref{eq:eta_turb}) and (\ref{eq:eta_OD_est}) shows that the Ohmic dissipation dominates over the turbulent diffusion, $\eta_{\rm OD}\gg \eta_{\rm t}$, inside the `dead' zone, $r\sim 1$~au, with low ionization fraction, $x\sim 10^{-15}$. The turbulent diffusivity could be of the order of Ohmic diffusivity in the innermost region of the disk, $r\lesssim 0.1$~au, where $T\gtrsim 1000$~K and $x\gtrsim 10^{-6}$. \citet{RRS1996} came to a similar conclusion and showed that the role of turbulent diffusion  depends strongly on Pr$_{\rm m}$. 
Generally speaking, the consideration of the turbulent diffusion requires to take $\alpha$-effect and the dynamo generation of the magnetic field into account.

\subsection{Angular momentum transport}
\label{sec:angmom}
\paragraph{Main equation.}
Let us consider the connection between the magnetic field and the angular momentum transport in disks. The equation of angular momentum evolution can be written as
\begin{eqnarray}
\frac{\partial I_z}{\partial t} + \frac{1}{r}\frac{\partial}{\partial r}\left(rv_rI_z\right) +  \frac{\partial}{\partial z}\left(v_zI_z\right) &=& \frac{1}{r}\frac{\partial}{\partial r}\left(r^2\sigma'_{r\varphi} + r^2\frac{B_rB_\varphi}{4\pi}\right) + \nonumber\\
& & \frac{\partial}{\partial z}\left(\frac{rB_zB_\varphi}{4\pi}\right),\label{eq:I}
\end{eqnarray}
where $I_z=\rho rv_\varphi$ is the angular momentum per unit volume, $\textbf{v}=\{v_r,\, v_\varphi,\, v_z\}$ and $\textbf{B}=\{B_r,\, B_\varphi,\,B_z\}$ are the magnetic field and velocity vectors in cylidrical coordinates, respectively. 

The second term on the left-hand-side of equation~(\ref{eq:I}) describes the advection of the angular momentum with the gas flow in the radial direction (accretion), the third term corresponds to the momentum flux in the vertical direction (outflows). 

The first term on the right-hand-side (under the sing of the radial derivative) corresponds to the torques acting on the gas in the disk plane, namely: 
turbulent stresses ($\propto \sigma'_{r\varphi}$)  and magnetic tensions ($\propto B_rB_\varphi/4\pi$).  
The term proportional to $B_rB_\varphi/4\pi$ in equation (\ref{eq:I}) comes from large-scale magnetic field tensions, the effect of which on the evolution of angular momentum in accretion disks have not yet been fully investigated.

The second term on the right-hand side describes vertical extraction of the angular momentum by the torque connected with large-scale magnetic tensions $B_zB_\varphi/4\pi$. This magnetic torque is tightly connected to the outflows from the surface of the disk~\citep[see, e.g., recent review by][]{lesur2021}.

\paragraph{Role of the turbulence.}
Turbulence in PPDs is probably of MHD nature, although purely hydrodynamic instabilities are not excluded as a source of the turbulence~\citep[see][]{lesur2023}. MHD turbulence can be generated due to the magnetorotational instability~\citep[MRI,][]{velikhov1959, chandrasekhar1960, bh1991}.
The MRI develops under conditions of $\beta > 3$ and $\lambda_{\rm MRI}<H$, where $\lambda_{\rm MRI} = 2\pi\eta/v_{\rm A}$ is the most unstable wavelength, $v_{\rm A}$ is the Alfv{\'e}n speed. Both local and global numerical simulations in the ideal MHD limit confirmed the prediction of the linear theory~\citep[e.g.,][]{stone1996, flock2013}. The turbulent redistribution of the angular momentum in the disk plane causes gas to accrete with speed
\begin{equation}
v_r \approx -\frac{3}{2}\frac{\nu_{\rm t}}{r}
\end{equation}
equivalent to mass accretion rate
\begin{equation}
\dot{M} = 3\pi\nu_{\rm t}\Sigma,
\end{equation}
where $\Sigma$ is the gas surface density.

Accretion and protoplanetary disks exhibit complex ionization structure due to non-uniform distribution of the ionization sources~\citep{gammie1996}. The innermost regions of the disk are well ionized due to the thermal ionization of alkali metals. Surface layers of the disk and its periphery are effectively ionized by cosmic rays, as well as by XR- and UV-radiation from the star. The attenuation of the external ionizing radiation leads to the formation of a region with a low ionization degree, $x\lesssim10^{-12}$, and efficient Ohmic dissipation near the midplane of the disk. 
As a consequence, this region is characterized by damped MHD turbulence~\citep{gressel2015} and reduced magnetic field strength~\citep[][see also previous paragraph]{dkh2014}. In absence of dust grains, the turbulent mixing can homogenize the gas in the vertical direction keeping the ionization fraction at the level sufficient for the activation of the MRI~\citep{inutsuka2005}.

Numerical simulations that take into account non-ideal MHD effects have shown that the MRI is also damped near the surface of the disk above the `dead' zone and partly at the disk periphery due to the effect of magnetic ambipolar diffusion~\citep{bai2013}. 

Therefore, MHD turbulence can be damped at least in some parts of PPDs. This conclusion is partially supported by recent measurements of molecular line widths in submm-wavelengths in several PPDs. The analysis of the turbulent broadening of the lines revealed turbulent speeds $v_{\rm  turb}< 0.15\,c_{\rm s}$, which corresponds to  $\alpha < 0.02$~\citep[see recent review by][]{rosotti2023}.

\paragraph{Role of the outflows.}
The theory of MHD outflows was developed by~\citet{bp1982} in application to the accretion disks of black holes and by~\citet{pudritz1986} in application to the disks in YSOs. The basic is that strengthening of the magnetic field near the surface of the disk due to differential rotation creates a magnetic torque (third term on the left-hand-side in (\ref{eq:I})), which transports angular momentum from the gas in the disk to the medium above the disk. Extraction of the angular momentum from the surface layers of the disk allows gas to accrete with speed
\begin{equation}
v_r \approx -\frac{B_\varphi^+B_z}{2\pi\Omega_{\rm k}\Sigma}
\end{equation}
equivalent to mass accretion rate
\begin{equation}
\dot{M} \approx \frac{r^2}{v_{\rm k}}B_\varphi^{+}B_z,\label{eq:Mdot_B}
\end{equation}
where $B_\varphi^+$ is the azimuthal component of the magnetic field near the surface of the disk, $v_{\rm k}=\Omega_{\rm k}r$ is the keplerian speed.

Gas near the surface of the disk can flow out of the disk if it can  escape the potential well determined by gravitational and centrifugal forces. In the case of `cold'  gas, for which thermal energy is much smaller than the kinetic energy, this corresponds to a condition that the poloidal magnetic field lines are inclined by more than 30 degrees to the axis of rotation~\citep[see, e.g., reviews by][]{spruit1996, lesur2021}. In the opposite case of a `hot' gas, which is heated efficiently by external radiation, this condition is less strict. \citet{bai2016} suggested the term `magneto-thermal wind' for the latter case. Further acceleration of the gas along the magnetic field lines is caused by the vertical magnetic pressure gradient, as in the mechanism proposed by~\citet{uchida1985}. The inertia of the gas above the Alfv{\'e}n point, where flow speed equals local Alfv{\'e}n speed, leads to the collimation of the outflow at further distances. The degree of collimation depends on the radial distribution of the poloidal magnetic field in the disk~\citep[see review by][]{pudritz2019}. 

Outflows are very efficient at extracting angular momentum from the disk in the sense that the specific angular momentum stored in the gas at the Alfven point $r_{\rm A}$ is $\left(r_{\rm A}/r_0\right)^2$ times larger than the specific angular momentum of the gas in the disk at a radial distance $r_0$ from the star. In the case, when the accretion is driven solely by the outflows with the outflow rate $\dot{M}_{\rm w}$, one have for the accretion rate $\dot{M}\approx \left(r_{\rm A}/r_0\right)^2 \dot{M}_{\rm w}$. Models of MHD outflows predict that $\left(r_{\rm A}/r_0\right)^2>3/2$, i.e. only a small fraction of gas is needed to be flow out of the disk in order to provide accretion.

Both local and global MHD simulations of PPDs found that the outflows form in a broad range of underlying disk conditions~\citep{bai2013, lesur2014, gressel2015, gressel2020}. Based on the predictions of the simulations and the observed universality of outflows in YSOs, it is now widely accepted that this mechanism may be as important as turbulence in the evolution of the angular momentum in PPDs. However, direct comparison of model predictions with observations is difficult due to insufficient knowledge of the large-scale magnetic field distribution in disks.

\paragraph{Comparison of angular momentum transport mechanisms.} Relative roles of different angular momentum transport mechanisms can be analysed using corresponding time scales. Let us estimate the time scales for the following limiting cases: a) `dead' zone, where the magnetic field is not strengthened due to the action of Ohmic and ambipolar diffusions, so that  $(B_r,\,B_\varphi)\ll B_z$ (see also discussion of Figure~\ref{fig:MHD} above); b) a region of efficient magnetic field generation, in which all three magnetic field components can be comparable  (i.e. outside the `dead' zone), $B_r\sim B_\varphi\sim B_z$.

Turbulent redistribution of the angular momentum takes place on a `viscous' time scale:
\begin{equation}
t_{\rm v} = \frac{r^2}{\nu_{\rm t}} \approx \alpha^{-1}\left(\frac{H}{r}\right)^2t_{\rm k},\label{eq:t_v}
\end{equation}
where $t_{\rm k}=\Omega_{\rm k}^{-1}$ is a dynamical time scale.

Magnetic tensions in the plane of the disk redistribute the angular momentum on a time scale
\begin{equation}
t_{Br} = \frac{4\pi I_{z}}{B_r B_\varphi} \approx\frac{1}{2}\beta\left(\frac{H}{r}\right)^{-2}t_{\rm k}.\label{eq:t_Br}
\end{equation}
Approximate equality in (\ref{eq:t_Br}) is derived under assumption of efficient magnetic field generation, $B_r\sim B_\varphi$. From (\ref{eq:t_v}) and (\ref{eq:t_Br}) we conclude that the magnetic tensions in the plane of the disk dominate over turbulent stresses outside the `dead' zone, if
\begin{equation}
\beta < 100 \left(\frac{\alpha}{0.01}\right)^{-1}.
\end{equation}
Relation between the radial magnetic tensions and turbulent stresses inside the `dead' zone, where the radial and azimuthal magnetic field components are not generated due to strong diffusion, depends on the values of corresponding magnetic Reynolds numbers.

Vertical magnetic tensions act on the time scale of
\begin{equation}
t_{Bz} \approx \frac{4\pi I_z}{B_\varphi B_z} \frac{H}{r} \approx \frac{1}{2}\beta\left(\frac{H}{r}\right)^{-1}t_{\rm k}.\label{eq:t_Bz}
\end{equation}
Second approximate equality in (\ref{eq:t_Bz}) is derived for the conditions outside the `dead zone', $B_\varphi \sim B_z$. Comparison of (\ref{eq:t_Br}) and (\ref{eq:t_Bz}) shows that
\begin{equation}
t_{Bz} \approx \frac{1}{2}\left(\frac{H}{r}\right)t_{Br},
\end{equation}
i.e. vertical magnetic tensions (causing outflows) dominate over radial ones, since $H\ll r$. On the contrary, $B_\varphi \ll B_z $ and radial tensions are dominant inside the `dead' zone.

\subsection{MHD models of the PPDs}
\label{sec:dynamics}

One of the major uncertainties in the theory of accretion and protoplanetary disks is the distribution of the large-scale magnetic flux threading the disk. The strength and geometry of the magnetic field depend on the interaction of induction amplification and diffusion. Accretion leads to the magnetic field dragging towards the star, i.e. strengthening of the vertical magnetic field component $B_z$ and generation of the radial component, $B_r$. This process  is opposed by the outward radial diffusion of $B_z$.  Furthermore, differential rotation generates $B_r$ and $B_\varphi$ from $B_z$. In the geometrically thin disk, $B_r$ and $B_\varphi$ diffuse mainly in the vertical direction.

\citet{lubow1994} formulated a simple 1D kinematic model of poloidal magnetic field advection/diffusion in turbulent accretion disks. They found that turbulent diffusion prevents significant magnetic field dragging in the accretion disks characterised by a realistic value of the turbulent magnetic Prandlt number of order unity. \citet{BK2007} and \citet{guilet2013} pointed out that the dragging can still be efficient because the turbulent diffusivity in the vertical direction is non-uniform, and some layers of the disk may not be turbulent.  \citet{guilet2014} have found stationary distribution of $B_z$ described by a self-similar power-law radial profile with the power-law index ranging from $0$ for efficient diffusion to $-2$ for efficient advection.  The latter value is consistent with the analytical solution of~\citet{okuzumi2014}, which was derived in a similar problem statement. The profile $B_z\propto r^{-2}$ is an expected result of magnetic flux freezing in the disk with an initially uniform magnetic field. \citet{shu2007} took into account that the tensions of poloidal magnetic field slow-down the gas rotation, and derived an analytical solution  $B_z~\propto r^{-11/8}$ for the stationary turbulent accretion disks with fixed degree of sub-keplerian rotation.

\citet{RRS1996} calculated the magnetic field in accretion disks using a kinematic approximation, taking turbulent diffusion and Ohmic dissipation into account. They pointed out that differential rotation can lead to the generation of a strong azimuthal magnetic field component. The authors also  found that a non-uniform radial profile of the ionization fraction and Ohmic diffusivity in viscous PPDs leads to non-monotonic radial profiles of magnetic field components with minimum magnetic field strength in the region of the ionization minimum.

Based on the theory of the fossil magnetic field, \citet{dkh2014} developed a MHD model of the accretion disks with a large-scale magnetic field~\citep[see also][]{kh2017}. The model is based on the approximations of~\citet{ss1973}. A stationary geometrically thin disk in magnetostatic and centrifugal equilibrium is considered. 
The thermal structure of the disk is simulated taking into account turbulent heating, as well as heating from the radiation of the central star and the interstellar CR. The magnetic field is calculated taking into account Ohmic dissipation, magnetic ambipolar diffusion, magnetic buoyancy and the Hall effect. The ionization structure of the disk is modelled taking into account radiative recombinations and recombinations on dust grains, ionization by CR, XR and radionuclides. 
The main parameters of the model are the mass accretion rate $\dot{M}$, the turbulence parameter $\alpha$, the characteristics of dust grains and ionizing radiation, and the opacity parameters. The model predicts that magnetic field scales with surface density, $B_z\propto \Sigma$, in the ideal MHD case. This solution reduces to the radial profile $B_z\propto r^{-2}$ in the particular case of a uniform disk. Simulations of the accretion disk structure using the model have shown that the combination of all relevant MHD effects results in non-uniform magnetic field distribution and complex magnetic field geometry in PPDs. The magnetic field is quasi-toroidal, $B_\varphi\gtrsim (B_r,\, B_z)$, in the innermost region, $r<0.3-0.5$~au, where thermal ionization operates. The runaway growth of the toroidal magnetic field is limited by the magnetic buoyancy in this region. Initial poloidal magnetic field geometry is preserved, $B_z\gg (B_r,\,B_\varphi)$, due to efficient Ohmic and ambipolar diffusion within the `dead' zone at $0.3<r<(10-20)$~au for typical parameters. The magnetic field can be quasi-radial, $B_r\sim B_z$, or quasi-toroidal at the periphery of the disk, $r>20$~au,  depending on  the dust grain parameters and ionization sources.

Long-term global multidimensional numerical simulations of PPDs evolution taking into account all MHD and ionization/recombinations effects, as well as radiation transfer, are still challenging due to the limitations on the time steps in corresponding numerical schemes. Local shearing-box approximation is often used to study the evolution of the PPDs taking into account Ohmic dissipation and magnetic ambipolar diffusion~\citep{simon2013}, as well as the Hall effect~\citep{lesur2014}. It is important to note that the local simulations cannot properly handle the vertical boundary conditions and to study the actual geometry of the magnetic field in the disk~\citep[e.g., see discussion in][]{lesur2021}.
Current global 2D and 3D simulations that consider extended regions of the disk have focused primarily  on the regions of efficient Ohmic and/or ambipolar diffusion. To facilitate the computations, simplified ionization-recombination models and/or prescribed diffusivities were implemented. Evidence has been found that the MHD effects can lead to the  formation of substructures in PPDs~\citep{bethune2017, suriano2018, riols2020}. \citet{cui2021} have also shown that both magnetically driven outflows and MHD turbulence can contribute to angular momentum transport in ambipolar diffusion dominated regions. The most sophisticated models to date  combined non-ideal MHD modelling with complex ionization models~\citep{bai2017} and radiative transfer calculations~\citep{wang2019, gressel2020}. It was found that the MHD outflows can be the main drivers of angular momentum transport under the considered parameters.

\paragraph{Fossil magnetic field strength in PPDs.}

\begin{figure}
\includegraphics[width=0.99\textwidth]{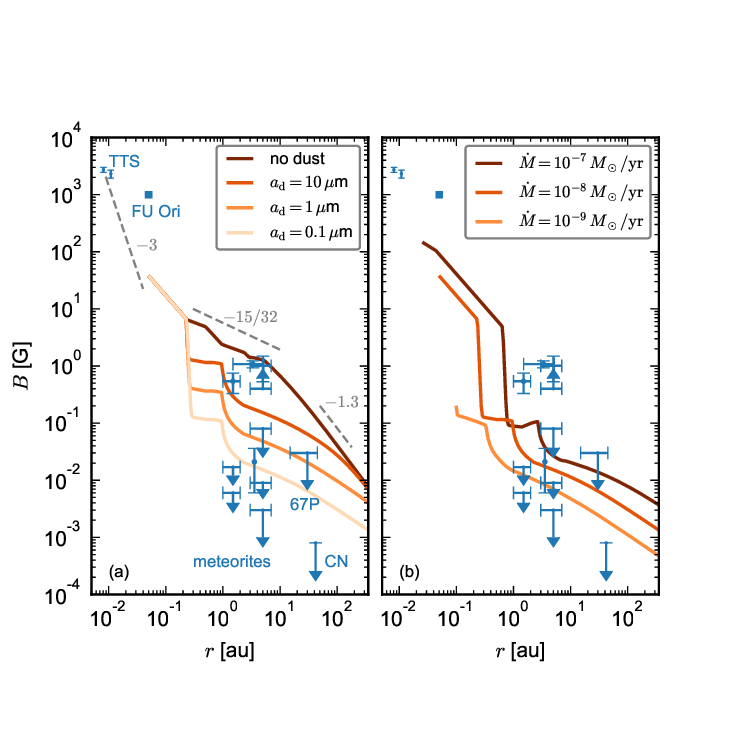}
\caption{Comparison of the observational constraints and theoretical predictions on the magnetic field strength in PPDs. Observational data are taken from table~\ref{Table:B}: `TTS': surface magnetic fields in T Tauri stars, `FU Ori': magnetic field strength in the FU Ori disk, `meteorites': estimates based on the measurements of the remnant magnetism of solar system meteorites, `67P': magnetic field of the comet 67P/Churyumov-Gerasimenko, `CN': upper limit on the magnetic field strength in the TW Hya disk based on non-detection of Zeeman splitting in CN lines. {\it Panel a}: radial profile of the magnetic field strength computed with the help of MHD model of~\citet{kh2017} for different dust grains sizes and fixed mass accretion rate $\dot{M}=10^{-8}\,M_\odot\,\mbox{yr}^{-1}$ (`no dust' corresponds to the case of absent dust grains). Dashed lines with labels depict characteristic slopes. {\it Panel b}: analogous profiles for different accretion rates and fixed dust grain size of $0.1\mu$m.}
\label{fig:B}
\end{figure}

Let us analyse the strength of the fossil magnetic field in the accretion disk of a typical solar-mass T Tauri star. 
In figure~\ref{fig:B}, we plot radial profiles of $B_z$ calculated using the model of~\citet{kh2017} for various dust grains radii and mass accretion rates. The turbulence parameter is fixed, $\alpha=0.01$. The ionization parameters correspond to the fiducial run in~\citet{kh2017}. Observed constraints on the magnetic field strength compiled in table~\ref{Table:B} are also shown for comparison.

Figure~\ref{fig:B}(a) shows that, in the case of absent dust grains, fossil magnetic field strength in the disk is maximum. According to our simulations, the ionization fraction does not fall below  $10^{-11}-10^{-10}$, the magnetic field is frozen in the gas, and its strength is proportional to the gas surface density, $B_{\rm f}\propto \Sigma$. Such a situation occurs also in the case of large dust grains, $a_{\rm d}\geq 1$~mm~\citep{khd22}.
In this case, magnetic field strength decreases from $\approx 50$~G near the inner edge of the disk, $r_{\rm in}=0.05$~au, to $3$~G at $r=3$~G, and down to $0.01$~G at the disk's outer edge, $r_{\rm out}=140$~au. The typical slope of the $B_{\rm f}(r)$ profile is $-3/8$ in the region $0.3<r<10$~au, and it is more steep, $B_{\rm f}\propto r^{-1.3}$, at the periphery of the disk. 

Recombinations of electrons on dust grains reduce the ionization fraction and consequently the magnetic field strength. Figure~\ref{fig:B}(a) shows that the magnetic field strength in the case of dust grains with an average size of $0.1\mu$m is 2 orders of magnitude lower than $B_{\rm f}$. In this case, the typical magnetic field strength is of $0.1$~G at $1$~au and of $10^{-3}$~au near the outer edge of the disk. The `flat' profile of $B(r)$ in the region $0.3-1$~au is explained by the evaporation of the icy mantles of the dust grains and corresponding decrease of the rate of recombinations on dust grains.

According to Figure~{\ref{fig:B}(a), the increase in dust grains size leads to an increase in the magnetic field strength. This can be illustrated by the following analytic solution derived by~\citet{dkh2014} taking magnetic ambipolar diffusion into account:
\begin{eqnarray}
B_z &=& 0.024\,\mbox{G} \left(\frac{\xi}{10^{-17}\,\mbox{s}^{-1}}\cdot \frac{a_{\rm d}}{0.1\,\mu\mbox{m}}\right)^{1/2}\left(\frac{\alpha}{0.01}\right)^{1/16}\cdot \nonumber\\
& & \left(\frac{\dot{M}}{10^{-8}\,M_{\odot}/\mbox{yr}}\right)^{3/9}\left(\frac{M}{1\,M_{\odot}}\right)^{5/32}\left(\frac{r}{1\,\mbox{au}}\right)^{-15/32}\label{Eq:Bz_mad},
\end{eqnarray}
where $\xi$ is the ionization rate. Solution (\ref{Eq:Bz_mad}) shows that $B\propto a_{\rm d}^{1/2}$, which is a consequence of the corresponding solution for the ionization fraction, $x\propto a_{\rm d}^{1/2}$. Observations and theoretical modelling show that the evolution of PPDs is characterized by the growth of dust grains up to $1$~mm in size~\citep[see review by][]{testi2014}. During evolution, large dust grains sediment toward the midplane of the disk and undergo radial drift toward the star. Analysis of the Figure~\ref{fig:B}(a) suggests that the evolution of dust grains sizes and distribution in the disk will lead to corresponding changes in the magnetic field strength. In particular, dust grain growth leads to an increase in the magnetic field strength.

The innermost region of the disk, $r<0.3$~au is characterized by a large ionization fraction caused by thermal ionization of alkali metals. In this case, the magnetic field profile is frozen in gas.

Figure~\ref{fig:B}(b) shows that it is not the shape of the radial profile $B(r)$ that changes with varying mass accretion rate, but rather the profile as a whole `shifts' towards the star with decreasing $\dot{M}$. This is because the disks with greater accretion rates are hotter and denser, and their inner boundary is closer to the star.  The increase in temperature increases the  innermost region of thermal ionization and frozen magnetic field. For example, this region occupies $r<1$~au in the case of $\dot{M}=10^{-7}\,M_\odot\,\mbox{yr}^{-1}$, while it is completely missing in the case of $\dot{M}=10^{-9}\,M_\odot\,\mbox{yr}^{-1}$. According to observations, the mass accretion rate decreases during the evolution of the PPD~\citep[see][]{hartmann2016}. Our results suggest that the fossil magnetic field strength also decreases over time.

Our simulations show that under typical parameters magnetic field strength at the inner edge of the disk, $r\approx 0.05$~au is $\gtrsim 100$~G, which is close the stellar dipole magnetic field at this distance. We interpreted this as a consequence of the common fossil nature of these fields. Estimates of the magnetic field strength based on remnant magnetisation of meteorites range widely between $\approx 2\cdot 10^{-3}$  and $1$~G in the region of terrestrial planets $r=1-7$~au. Such scattering could be a reflection of the diversity and/or evolution of conditions in the protosolar nebula. This could be particularly related to the changes in accretion rate (and therefore age) and dust grain radius.

Another approach to analyse the magnetic field in PPDs is to estimate the magnetic field strength required to provide observed accretion rates by magnetically driven outflows~\citep[e.g., see][]{wardle2007}. The expression~(\ref{eq:Mdot_B}) gives $B_\varphi B_z=\dot{M}v_{\rm k}/r^2$, which corresponds to the lower estimate for the magnetic field strength
\begin{equation}
B\geq 0.09\left(\frac{\dot{M}}{10^{-8}\,M_\odot\,\mbox{yr}^{-1}}\right)^{1/2}\left(\frac{M}{1\,M_\odot}\right)^{1/4}\left(\frac{r}{1\,\mbox{au}}\right)^{-5/4}\,\mbox{G}.\label{eq:B_Mdot}
\end{equation}
Solution (\ref{Eq:Bz_mad}) and estimate (\ref{eq:B_Mdot}) give similar values close to $r\sim 1$~au, proving consistency of different approaches and indicating the comparable role of turbulence and outflows in terms of the efficiency of angular momentum transport. 

\paragraph{Dynamical influence of the magnetic field on the structure of the disk.}

The MHD models discussed above predict that the magnetic field strength varies significantly across the disk and the corresponding value of plasma beta for typical parameters is the in range $10^2-10^5$. Such a magnetic field is kinematic, i.e. it has no influence on the structure of the disk. Nevertheless, even such a small magnetic field plays a key role in the disks' dynamics, since the development of the MRI is only possible in regions with $\beta>3$. 

\citet{khd22} have shown that in the case of large dust grains, $a_{\rm d}\geq 1$~mm, and increased ionization rate, $\dot{M}\ge 10^{-7}\,M_\odot\,\mbox{yr}^{-1}$, a dynamically strong magnetic field, $\beta\sim 1$, can be generated in the innermost regions of the disk and at its periphery.  In such a case, magnetic tensions slow down the gas rotation speed by $0.5-1$~\% compared to the keplerian speed. This `magnetic' deceleration is greater than or on order of the hydrodynamic deceleration caused by the radial gas pressure gradient. The model predictions are consistent with modern observations of several PPDs, which indicate the sub-keplerian rotation in the outer regions of the disks with a maximum deviation of up to several percent of $v_{\rm k}$~\citep{pinte2018, dullemond2020, teague2021}.

Sub-keplerian rotation is usually considered to be the cause of the radial drift of dust grains and small bodies in PPDs~\citep{w77}. We conclude that the speed of radial drift increases in the region of a dynamically strong magnetic field, which could influence the process of formation and evolution of planetesimals. 

The magnetic field can be dynamically strong in the surface layers of the disk, where the gas pressure is low. As a result, the magnetic pressure gradient affects the vertical structure of the disk, reducing or increasing the thickness of the disk depending on the surface conditions~\citep{ogilvie2001, lizano2016, khd22}.

The thermal structure of the disk can also be influenced by MHD effects. In the region of large currents, Ohmic and ambipolar heating  lead to additional heating of the gas. Following the approach of~\citet{shu2007}, \citet{lizano2016} found that Ohmic heating dominate over viscous heating in disks with large accretion rates, such as in FU Ori. \citet{mori2019} performed local MHD simulations of the innermost region of a PPD for fixed $\beta=10^2-10^6$, low gas opacity, $\kappa<0.5$~cm$^2$~g$^{-1}$, and found that the Ohmic heating affects the gas temperature only in the surface layers of the disk. \citet{khd2019mhd} have shown that Ohmic and ambipolar heating increase the temperature of the disk near the borders of the `dead' zone by $10-100$~K. \citet{bethune2020} concluded that these MHD effects influence the temperature of the disk at
$r=0.5-4$~au in the case of sufficiently high gas opacity, $\kappa>0.1$~cm$^2$~g$^{-1}$.
 
\section{Summary}
\label{sec:summary}

Analysis of modern observations shows that the accretion and protoplanetary disks of young stars have large-scale magnetic field. The magnetic field plays an important role in the evolution of PPDs, and its influence is largely determined by magnetic field strength and  geometry. The observational data obtained so far are scanty and mostly indirect, therefore theoretical MHD models of PPDs are of great importance. 

Key predictions of the MHD models of PPDs are following.
\begin{itemize}
\item MHD effects cause complex geometry of the magnetic field and significant variations in its strength across the disk. 
Contemporary observational data are consistent with the predictions of the fossil magnetic field theory, although they are scanty and largely indirect.

\item The magnetic field can be dynamically strong affecting the vertical structure of the disk and the rotation speed of the gas. The Ohmic and ambipolar diffusion can increase the gas temperature. Above  dynamical effects can lead to observably verifiable consequences: variations in the disk thickness, increase in the radial drift speed of dust grains and even planets, stabilization of gravitational instability, etc.
\end{itemize}

The following problems and perspectives in both observations and theory can be outlined to date.
\begin{itemize}
\item It is a significant and challenging task to separate the roles played by the magnetic field, radiation scattering, and geometrical effects in the polarization of the disk's thermal emission~\citep{ohashi2018, lee2018}. \citet{delangen2023} found that near-infrared circular polarization can also be useful for studying the magnetic field structure and dust properties at the disk surface.
\item Zeeman splitting measurements of the CN molecular lines are a perspective instrument for studying the magnetic field strength in PPDs~\citep{kh21chem, lankhaar2023}.
\item Models of PPDs has to take into account evolution of angular momentum in the disk taking into account the combination of turbulence and the tensions of a large-scale magnetic field in different regions of the disk. Empirical models of the disks with magnetically driven outflows have recently been presented~\citep{suzuki2016, tabone2022}. The models use a parametrization of the outflow properties similar to the $\alpha$-parametrization of turbulence. Such an approach makes it possible to study the long-term evolution of PPDs and predict observable differences between the disks characterized by turbulence or outflow-driven angular momentum transport.
\item Although there is evidence that the hydromagnetic dynamo operates in disks~\citep{hawley1996, gressel2015b}, its role in magnetic flux transport in PPDs is not yet fully understood. So far, the magnetic field generation by $(\alpha-\Omega)$-dynamo has only been studied in the kinematic approximation~\citep{moss2016}. In this context, the relationship between the small-scale turbulent and large-scale magnetic fields in PPDs requires further investigation.
\item The evolution of the magnetic flux of the disk, its dissipation at later stages of disk evolution (transition and debris disks), magnetic fields of protoplanets are open questions.
\end{itemize}

{\bf Acknowledgements.} The study of the magnetic fields in the outflows from YSOs (section~\ref{sec:obs}) is financially supported by Russian Science Foundation (project 23-12-00258). The study of the fossil magnetic field in PPDs (section~\ref{sec:dynamics}) is supported by the Theoretical Physics and Mathematics Advancement Foundation `BASIS' (project~23-1-3-57-1). The author thanks professor Andrey Sobolev for consultations regarding masers in star formation regions. The author is thankful to the reviewer for giving useful comments, which helped to improve the quality of the review.

\end{document}